\documentclass[table]{ccjnl}
\usepackage{lipsum,amsmath}
\usepackage{cuted}
\usepackage{bm}
\graphicspath{{figures/}}

\title{Wideband Channel Estimation for THz Massive MIMO}
\author{Jingbo Tan\inst{1} and Linglong Dai\inst{1}}

\receiveddate{Jan. 25, 2021}
\reviseddate{Mar. 14, 2021}
\Editor{Zhi Chen}

\address[1]{Beijing National Research Center for Information Science and Technology (BNRist) as well as the Department of Electronic Engineering, Tsinghua University, Beijing 100084, P. R. China}

\begin{document}

\maketitle

\begin{abstract}
Terahertz (THz) communication is considered to be a promising technology for future 6G network. To overcome the severe attenuation and relieve the high power consumption, massive multiple-input multiple-output (MIMO) with hybrid precoding has been widely considered for THz communication. However, accurate wideband channel estimation, which is essential for hybrid precoding, is challenging in THz massive MIMO systems. The existing wideband channel estimation schemes based on the ideal assumption of common sparse channel support will suffer from a severe performance loss due to the beam split effect. In this paper, we propose a beam split pattern detection based channel estimation scheme to realize reliable wideband channel estimation in THz massive MIMO systems. Specifically, a comprehensive analysis on the angle-domain sparse structure of the wideband channel is provided by considering the beam split effect. Based on the analysis, we define a series of index sets called as beam split patterns, which are proved to have a one-to-one match to different physical channel directions. Inspired by this one-to-one match, we propose to estimate the physical channel direction by exploiting beam split patterns at first. Then, the sparse channel supports at different subcarriers can be obtained by utilizing a support detection window. This support detection window is generated by expanding the beam split pattern which is determined by the obtained physical channel direction. The above estimation procedure will be repeated path by path until all path components are estimated. Finally, the wideband channel can be recovered by calculating the elements on the total sparse channel support at all subcarriers. The proposed scheme exploits the wideband channel property implied by the beam split effect, i.e., beam split pattern, which can significantly improve the channel estimation accuracy. Simulation results show that the proposed scheme is able to achieve higher accuracy than existing schemes.
\keywords{THz communication; massive MIMO; hybrid precoding; beam split; wideband channel estimation}
\end{abstract}
\section{Introduction}
Terahertz (THz) communication has been considered as one of the promising techniques for future 6G network, since it can provide tenfold bandwidth increase and thus support ultra-high transmission rate \cite{1,2,3,4,5}. To overcome the severe attenuation in the THz band (i.e., $0.1-10$ THz \cite{1}), massive multiple-input multiple-output (MIMO), which can generate directional beams by a large-scale antenna array, is essential for THz communication\cite{4}. However, the traditional fully-digital structure, where each antenna is connected to one radio-frequency (RF) chain, will introduce very high power consumption \cite{6}. To solve this problem, hybrid precoding structure can be used for THz communication \cite{7,8,9}, where the high-dimensional precoder is decomposed into a high-dimensional analog beamformer (usually realized by analog components \cite{10}) and a low-dimensional digital precoder (usually realized by a reduced number of RF chains). Thanks to the sparsity of THz channels, it has been proved that hybrid precoding is able to achieve the near-optimal achievable rate performance \cite{7,8,9}.

\subsection{Prior works}
To design an efficient hybrid precoder, the high-dimensional channel is essential at the base station (BS). However, channel estimation is challenging in massive MIMO systems with hybrid precoding structure \cite{11}. Specifically, since the number of RF chains is much smaller than the number of antennas in the hybrid precoding structure, the BS cannot obtain signals at each antenna element simultaneously. As a result, to obtain sufficient observation to accurately estimate the high-dimensional channel, the channel estimation overhead of conventional channel estimation scheme, e.g., least square (LS) scheme, will be unacceptable when the number of antennas is very large \cite{12}.

To deal with this problem, exploiting channel sparsity with the help of compressive sensing algorithms for channel estimation has been widely investigated to realize low-overhead channel estimation in massive MIMO systems \cite{13,14,15,16,17,18,19,20}. For example, a distributed compressive sensing based multi-user channel estimation scheme was proposed in \cite{13}, where the joint angle-domain channel sparsity among different users was utilized. \cite{14} proposed an orthogonal matching pursuit (OMP) based channel estimation scheme for massive MIMO systems with hybrid precoding structure by using the angle-domain channel sparsity. Besides, a joint channel estimation and tracking scheme was also proposed based on the framework of compressive sensing in \cite{15}. In addition, the channel estimation problem in lens-array based massive MIMO with a simple antenna switching network is investigated in \cite{16}, where a redundant dictionary and the corresponding compressive sensing based scheme are proposed.

However, these schemes in \cite{13,14,15,16} were designed for narrowband systems. Although these narrowband schemes can be extended in wideband systems, carrying out narrowband schemes subcarrier by subcarrier will result in high complexity due to a large number of subcarriers, especially in wideband THz massive MIMO systems. To realize efficient wideband channel estimation, wideband channel estimation schemes have been proposed for millimeter-wave massive MIMO systems \cite{18,19}. In particular, \cite{18} proposed a simultaneous orthogonal matching pursuit (SOMP) based scheme, where channels at different subcarriers were jointly estimated based on the assumption of common sparse channel support (i.e., the sparse channel supports at different subcarriers are the same). Besides, an OMP based wideband channel estimation scheme was proposed in \cite{19}, where the sparse channel supports at some subcarriers were independently estimated using the classical OMP algorithm, and then the wideband channel was recovered based on the common sparse channel support created by the already obtained sparse channel supports. Furthermore, \cite{20} proposed a close-loop sparse channel estimation solution for multi-user massive MIMO systems. Unfortunately, the ideal assumption of common sparse channel support in the above two schemes is not practical for THz systems due to the beam split effect \cite{21}. Specifically, the beam split effect can be seen as a serious situation of the widely known beam squint \cite{22}. It means because of the wide bandwidth and a large number of antennas in THz massive MIMO systems, the spatial channel directions at different subcarriers becomes separated from each other in the angle-domain, i.e, locate at different angle-domain samples. The beam split effect will induce frequency-dependent sparse channel supports at different subcarriers. Consequently, the assumption of common sparse channel support does not hold, which means the existing schemes for millimeter-wave massive MIMO \cite{18,19} will suffer from severe performance degradation in wideband THz massive MIMO systems. Although several channel estimation schemes for THz massive MIMO has been recently proposed, such as the low-rank matrix reconstruction based scheme \cite{23} and the joint activity detection and channel estimation scheme \cite{24}, they have not considered the frequency-dependent sparse channel support either. Hence, to the best of our knowledge, the wideband channel estimation in THz massive MIMO systems has not been well addressed in the literature.

\subsection{Our contributions}
In this paper, we propose an accurate beam split pattern detection based wideband channel estimation scheme in THz massive MIMO systems. The specific contributions of this paper can be summarized as follows.

\begin{itemize}
	\item We first analyze the angle-domain sparse structure of the wideband THz channel by considering the beam split effect. We prove that a series of index sets have the one-to-one match to different physical channel directions. These index sets are defined as beam split patterns, each of which is corresponding to a specific physical channel direction. By utilizing the one-to-one match between the physical channel direction and the beam split pattern, the physical channel direction can be accurately estimated.
	\item Based on the proof above,  we propose a beam split pattern detection based wideband channel estimation scheme. For each channel path component, the physical channel direction is firstly estimated by exploiting the beam split pattern. Then, the sparse channel supports at different subcarriers are determined by using a support detection window. This support detection window is generated by expanding the beam split pattern, which is corresponding to the already obtained physical channel direction. The above procedure will be repeated path by path until all path components are considered. Finally, the wideband channel can be recovered by only calculating elements on the total sparse channel support containing sparse channel supports for different path components. Thanks to the one-to-one match between the physical channel direction and the beam split pattern, the proposed scheme can precisely estimate the physical channel directions and corresponding sparse channel supports.
	\item The physical channel direction estimation accuracy of the proposed scheme is analyzed, and it shows that the physical channel direction can be precisely estimated with a probability approaching $1$. Extensive simulation results verify this analysis, and illustrate that the proposed beam split pattern detection based wideband channel estimation scheme can realize more accurate channel estimation than existing schemes.
\end{itemize}

\subsection{Organization and notation}

The remainder of this paper is organized as follows. In Section \ref{Sys}, the system model of a multi-user wideband THz massive MIMO system is introduced, and the channel estimation problem in this system is then formulated. In Section \ref{BSPD}, we first define the beam split pattern and prove the one-to-one match between the physical channel direction and the beam split pattern. Then, a beam split pattern based wideband channel estimation scheme is proposed, together with the corresponding performance and complexity analysis. Section \ref{Sim} illustrates the simulation results (Simulation codes are provided to reproduce the results presented in this paper: http://oa.ee.tsinghua.edu.cn/dailinglong/publications\\/publications.html.). Finally, conclusions are drawn in Section \ref{Con}.

\emph{Notation:} $(\cdot)^{T}$, $(\cdot)^{H}$, $(\cdot)^{\dagger}$, $\|\cdot\|_\mathrm{F}$, and $\|\cdot\|_{k}$ denote the transpose, conjugate transpose, pseudo-inverse, Frobenius norm, and $k$-norm of a matrix, respectively; $|\cdot|$ denotes
the absolute operator; $\mathbf{H}(i,j)$ denotes the element of the matrix $\mathbf{H}$ at the $i$-th row and the $j$-th column;  If set $\Xi=\cup_{i}\{(a_{i},b_{i})\}$, $\mathbf{H}(\Xi)$ denotes the vector composed of elements $\mathbf{H}(a_{i},b_{i})$; $\mathbf{I}_{N}$ represents the identity matrix of size $N\times N$.

\section{System Model}\label{Sys}
In this paper, we consider an uplink time division duplexing (TDD) based multi-user wideband THz MIMO system with orthogonal frequency division multiplexing (OFDM). The hybrid precoding structure is employed at the BS to reduce energy consumption, as shown in Fig. 1. The BS equips an $N$-antenna uniform linear array (ULA) \cite{25}, and utilizes $N_\mathrm{RF}$ RF chains to serve $K$ single-antenna users simultaneously with $M$ subcarriers. 

\begin{figure}
	\centering
	\includegraphics[width=0.49\textwidth]{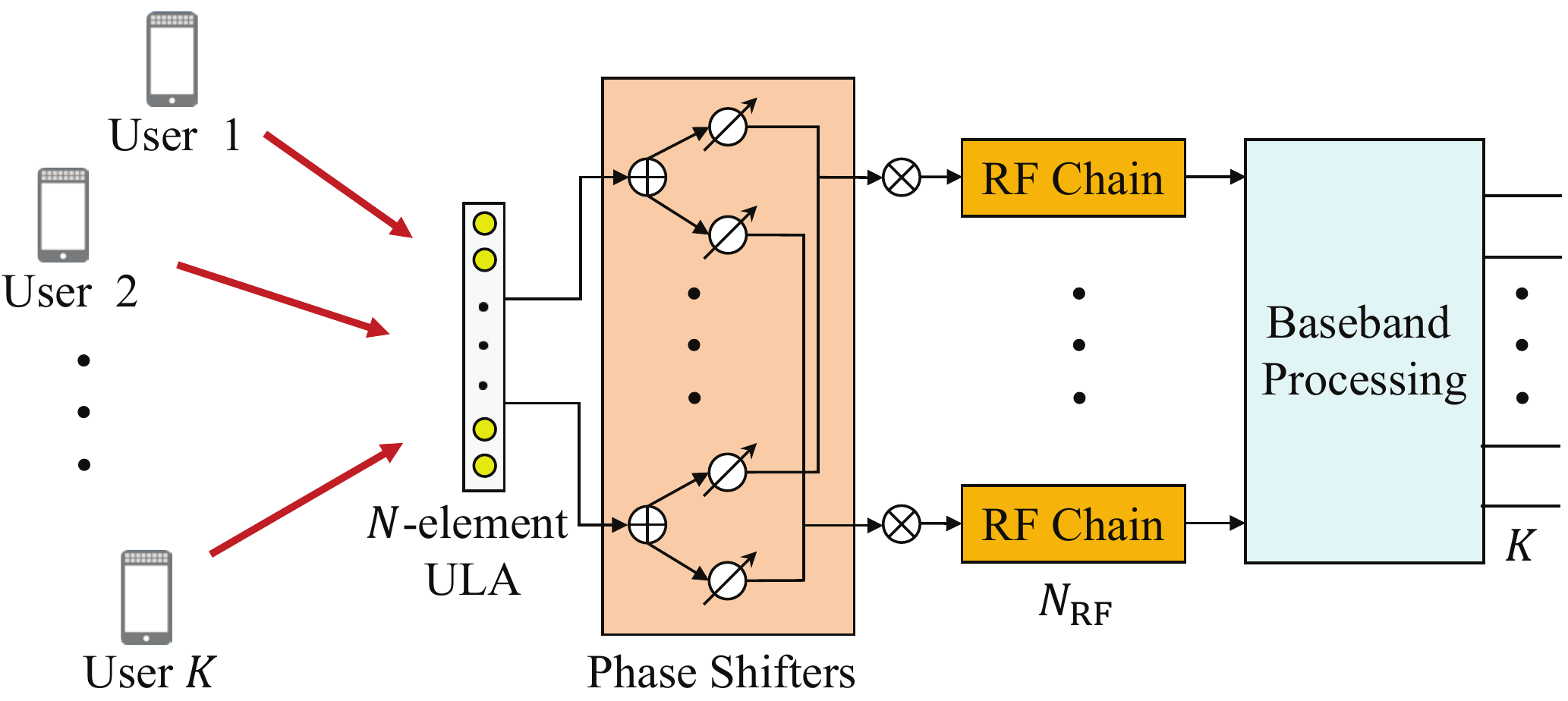}
	\caption{Wideband THz massive MIMO system with hybrid precoding.}
\end{figure}

\subsection{Channel model}
We adopt the widely used Saleh-Valenzuela multipath channel model \cite{26} in this paper. The channel $\mathbf{h}_{m}\in\mathcal{C}^{N\times 1}$ between the BS and a specific user  at the $m$-th subcarrer ($m=1,2,\cdots,M$) can be denoted as
\begin{equation}\label{1}
\mathbf{h}_{m}=\sqrt{\frac{N}{L}}\sum_{l=1}^{L}g_{l}e^{-j2\pi\tau_{l}f_{m}}\mathbf{a}(\theta_{l,m}),
\end{equation}
where $L$, $g_{l}$, and $\tau_{l}$ are the number of paths, the complex path gain of the $l$-th path, and the time delay of the $l$-th path, respectively, $\theta_{l,m}$ is the spatial channel direction of the $l$-th path at the $m$-th subcarrier, and $\mathbf{a}(\theta_{l,m})$ is the steering vector of $\theta_{l,m}$ with the following form
\begin{equation}\label{1-2}
\mathbf{a}(\theta_{l,m})=\frac{1}{\sqrt{N}}[1,e^{-j\pi\theta_{l,m}},e^{-j\pi2\theta_{l,m}},\cdots,e^{-j\pi(N-1)\theta_{l,m}}]^{T}. 
\end{equation}
Futhermore, the spatial channel direction $\theta_{l,m}$ can be represented as \cite{26}
\begin{equation}\label{2}
\theta_{l,m}=\frac{2f_{m}}{c}d\sin{\bar{\psi}_{l}}=\frac{2f_{m}}{c}d\psi_{l},
\end{equation}
where $\bar{\psi}_{l}$ is the physical channel direction of the $l$-th path with $\bar{\psi}_{l}\in(-\frac{\pi}{2},\frac{\pi}{2})$, $c$ denotes the light speed, $d$ is the antenna spacing usually set as $d=c/2f_{c}$ with $f_{c}$ representing the central frequency, $f_{m}$ is the frequency of the $m$-th subcarrier as $f_{m}=f_{c}+\frac{B}{M}(m-1-\frac{M-1}{2})$ with $B$ being the bandwidth. Without loss of generality, we define $\psi_{l}=\sin{\bar{\psi}_{l}}$ in (\ref{2}) as the physical channel direction in this paper for expression simplicity. We can observe from (\ref{2}) that unlike narrowband systems where the spatial channel direction $\theta_{l,m}\approx\psi_{l}$ is frequency-independent with $f_{m}\approx f_{c}$, in wideband systems, the spatial channel direction $\theta_{l,m}$ is frequency-dependent due to $f_{m}\neq f_{c}$. More seriously,  due to the large bandwidth, the spatial channel directions $\theta_{l,m}$ for subcarriers $m=1,2,\cdots,M$ will be quite different with a large gap between each other in THz massive MIMO systems. This effect, called as the beam split effect \cite{21}, will result in a serious performance loss for existing channel estimation schemes. 

The channel $\mathbf{h}_{m}$ can be transformed to its angle-domain representation by a spatial discrete Fourier transform matrix $\mathbf{F}\in\mathcal{C}^{N\times N}$. $\mathbf{F}$ contains $N$ orthogonal steering vectors covering the whole angle-domain as $\mathbf{F}=[\mathbf{a}(\bar{\theta}_{1}),\mathbf{a}(\bar{\theta}_{2}),\cdots,\mathbf{a}(\bar{\theta}_{N})]^{H}$, with $\bar{\theta}_{n}=\frac{2n-N-1}{N}, n=1,2,\cdots,N$. These physical channel directions $\bar{\theta}_{n}, n=1,2,\cdots,N$ can be seen as the angle-domain samples of the channel physical channel direction $\psi_{l}$. Correspondingly, the angle-domain channel $\bar{\mathbf{h}}_{m}\in\mathcal{C}^{N\times 1}$ can be denoted as
\begin{equation}\label{3}
\bar{\mathbf{h}}_{m}=\mathbf{F}\mathbf{h}_{m}=\sqrt{\frac{N}{L}}\sum_{l=1}^{L}g_{l}e^{-j2\pi\tau_{l}f_{m}}\mathbf{q}_{l,m},
\end{equation}
where $\mathbf{q}_{l,m}$ denotes the angle-domain representation of the $l$-th path component as
\begin{equation}\label{3-1}
\mathbf{q}_{l,m}=\mathbf{F}\mathbf{a}(\theta_{l,m})=[\Gamma(\theta_{l,m}-\bar{\theta}_{1}),\cdots,\Gamma(\theta_{l,m}-\bar{\theta}_{N})]^{T},
\end{equation}
with $\Gamma(x)=\frac{\sin{N\pi x/2}}{\sin{\pi x/2}}$ representing the Dirichlet Sinc function \cite{27}. Because of the power-focusing characteristic of the Dirichlet sinc function $\Gamma(x)$, the power of $\mathbf{q}_{l,m}$ focuses on a small number of elements decided by the spatial channel direction $\theta_{l,m}$. Additionally, since the number of scatters is limited in THz band, the number of path $L$ is usually quite small (e.g., $L=3$ \cite{28}). Therefore, we can conclude that the angle-domain channel $\bar{\mathbf{h}}_{m}$ is a sparse vector, and the sparse support of the angle-domain channel $\bar{\mathbf{h}}_{m}$ is decided by spatial channel directions $\theta_{l,m}$ for $m=1,2,\cdots,M$.

\begin{figure*}
	\centering
	\includegraphics[width=0.97\textwidth]{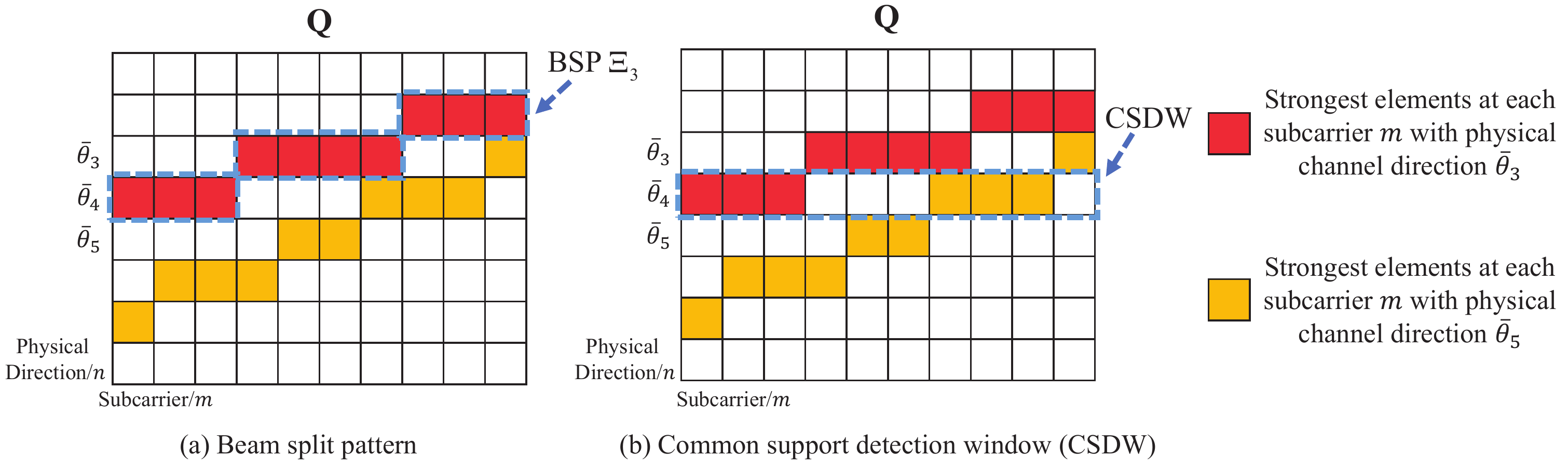}
	\vspace{-2mm}
	\caption{Illustration of the physical channel direction estimation with $\psi_{1}=\bar{\theta}_{3},\psi_{2}=\bar{\theta}_{5}$, where $\mathbf{Q}=\mathbf{Q}_{3}+\mathbf{Q}_{5}$: (a) Correct physical channel direction $\psi_{1}=\bar{\theta}_{3}$ can be estimated, since the BSP $\Xi_{3}$ can exactly capture the channel power incurred by the physical channel direction $\bar{\theta}_{3}$; (b) The existing assumption of a common sparse support will result in an incorrect estimate $\psi_{1}=\bar{\theta}_{4}$, because the highest power is captured at $\bar{\theta}_{4}$ when common support detection window is utilized to estimate the physical channel direction.}
	\vspace{-2mm}
\end{figure*}

\subsection{Problem formulation}\label{IIB}
In TDD systems, uplink channel estimation is carried out at the BS based on the received pilots transmitted by users. As orthogonal pilots are widely used \cite{26}, we can consider an arbitrary user without loss of generality for uplink channel estimation. By utilizing the sparsity of the angle-domain channel, the wideband channel estimation problem can be formulated as a joint sparse recovery problem. 

Specifically, we denote $s_{m,p}$ as the transmitted pilot at the $m$-th subcarrier in the time slot $p$. Then, the received pilots $\mathbf{y}_{m,p}\in\mathcal{C}^{N_\mathrm{RF}\times 1}$ at the $m$-th subcarrier is
\begin{equation}\label{4}
\begin{aligned}
\mathbf{y}_{m,p}&=\mathbf{A}_{p}\mathbf{h}_{m}s_{m,p}+\mathbf{A}_{p}\mathbf{n}_{m,p}\\
&\overset{(a)}{=}\mathbf{A}_{p}\mathbf{F}^{H}\bar{\mathbf{h}}_{m}s_{m,p}+\mathbf{A}_{p}\mathbf{n}_{m,p},
\end{aligned}
\end{equation}
where (a) comes from (\ref{3}), $\mathbf{A}_{p}\in\mathcal{C}^{N_\mathrm{RF}\times N}$ is the frequency-independent analog combining matrix satisfying $|\mathbf{A}_{p}(i,j)|=\frac{1}{\sqrt{N}}$ due to the hardware restriction (e.g., realized by phase-shifters \cite{29}), and $\mathbf{n}_{m,p}\in\mathcal{C}^{N\times 1}$ denotes the noise following the distribution $\mathbf{n}_{m,p}\sim\mathcal{CN}(0,\sigma^{2}\mathbf{I}_{N})$ with $\sigma^{2}$ being the noise power. Define $P$ as the length of transmitted pilots and assume $s_{m,p}=1$ for $p=1,2,\cdots,P$. Thus, the overall received pilots at the $m$-th subcarrier $\bar{\mathbf{y}}_{m}=[\mathbf{y}_{m,1}^{T},\mathbf{y}_{m,2}^{T},\cdots,\mathbf{y}_{m,P}^{T}]^{T}\in\mathcal{C}^{N_\mathrm{RF}P\times 1}$ can be denoted as
\begin{equation}\label{5}
\bar{\mathbf{y}}_{m}=\bar{\mathbf{A}}\bar{\mathbf{h}}_{m}+\bar{\mathbf{n}}_{m},
\end{equation}
where $\bar{\mathbf{A}}=[\mathbf{A}_{1}^{T},\mathbf{A}_{2}^{T},\cdots,\mathbf{A}_{P}^{T}]^{T}\in\mathcal{C}^{N_\mathrm{RF}P\times N}$ denotes the overall observation matrix, and $\bar{\mathbf{n}}_{m}$ is the effective noise satisfying $\bar{\mathbf{n}}_{m}=[\mathbf{n}_{m,1}^{T}\mathbf{A}_{1}^{T},\mathbf{n}_{m,2}^{T}\mathbf{A}_{2}^{T},\cdots,\mathbf{n}_{m,P}^{T}\mathbf{A}_{P}^{T}]^{T}\in\mathcal{C}^{N_\mathrm{RF}P\times1}$. Generally, as a result of channel correlation, channels at different subcarrier frequencies can be estimated jointly. Therefore, we rewrite (\ref{5}) as
\begin{equation}\label{15}
\mathbf{Y}=\bar{\mathbf{A}}\mathbf{H}+\mathbf{N},
\end{equation}
where we have $\mathbf{Y}=[\bar{\mathbf{y}}_{1},\bar{\mathbf{y}}_{2},\cdots,\bar{\mathbf{y}}_{M}]$, $\mathbf{H}=[\bar{\mathbf{h}}_{1},\bar{\mathbf{h}}_{2},\cdots,\bar{\mathbf{h}}_{M}]$, and $\mathbf{N}=[\bar{\mathbf{n}}_{1},\bar{\mathbf{n}}_{2},\cdots,\bar{\mathbf{n}}_{M}]$.

We can observe from (\ref{15}) that the wideband channel estimation problem is formulated as a joint sparse recovery problem, where the target is to recover the wideband sparse angle-domain channel $\mathbf{H}$ based on the observation matrix $\bar{\mathbf{A}}$ and the received pilots $\bar{\mathbf{y}}_{m}$. The sparsity of the channel $\mathbf{H}$ makes compressive sensing algorithms efficient with a significantly reduced length of pilots, i.e., $N_\mathrm{RF}P\ll N$ \cite{18}. Moreover, to further reduce the length of pilots and improve the channel estimation accuracy, existing wideband channel estimation schemes make use of the channel correlation by assuming channels at different subcarrier frequencies have a common sparse channel support \cite{18,19}. However, the ideal assumption of common sparse channel support may not hold in THz massive MIMO systems. Specifically, since the spatial channel direction $\theta_{l,m}$ in (2) is frequency-dependent due to the beam split effect \cite{21}, the sparse supports of channels $\bar{\mathbf{h}}_{m}$ at different subcarrier frequencies  for $m=1,2,\cdots,M$ should be quite different. This frequency-dependent sparse channel support will result in a severe performance loss for existing channel estimation schemes \cite{18,19}. Consequently, an accurate wideband channel estimation scheme is essential for THz massive MIMO systems.

\section{Beam Split Pattern Detection Based Channel Estimation}\label{BSPD}
In this section, we first define the beam split pattern based on the channel sparse structure, and reveal the one-to-one match between the physical channel direction and the beam split pattern. Then, by utilizing the one-to-one match, we propose the beam split pattern detection based channel estimation scheme to improve the estimation accuracy. Finally, the estimation accuracy analysis and complexity analysis of the proposed scheme are provided.

\subsection{Beam split pattern of wideband THz channel}
As described in Subsection \ref{IIB}, the beam split effect induces distinct spatial channel directions $\theta_{l,m}$, i.e., different sparse supports of $\bar{\mathbf{h}}_{m}$, at different subcarriers. Hence, it is hard to detect sparse channel supports accurately by using a common support detection window (SDW) at all subcarriers, like in existing schemes \cite{18,19}. To solve this problem, we reveal the one-to-one match between the physical channel direction $\psi_{l}$ and a specific indexes set, which is defined as the beam split pattern (BSP). Specifically, the BSP contains the element indexes of the angle-domain channel with the largest power at each subcarrier. The following \textbf{Lemma 1} provides a specific definition of BSP, and proves the one-to-one match between the physical channel direction and the BSP.
\newtheorem{thm}{Lemma}
\begin{thm} \label{lemma1}
	Considering an arbitrary path component with the physical channel direction $\psi_{l}$, we define $\mathbf{Q}_{n}=[\mathbf{q}_{l,1},\mathbf{q}_{l,2},\cdots,\mathbf{q}_{l,M}]$ and $n_{1}\in\{1,2,\cdots,N\}$. When we assume the physical channel direction $\psi_{l}$ locates on the angle-domain samples $\bar{\theta}_{n}$ with $\psi_{l}=\bar{\theta}_{n}$, the BSP $\Xi_{n}$ defined as 
	\begin{equation}\label{6}
	\Xi_{n}=\bigcup_{m=1}^{M}\left\{\left(\arg\min_{n_{1}}|\frac{f_{m}}{f_{c}}\bar{\theta}_{n}-\bar{\theta}_{n_{1}}|,m\right)\right\}
	\end{equation}
	has a one-to-one match to the physical channel direction $\bar{\theta}_{n}$.
\end{thm}
\emph{Proof:} According to the analysis in \cite{30}, the element index $n_{l,m}^\mathrm{max}$ of the angle-domain representation $\mathbf{q}_{l,m}$ with the largest power is
\begin{equation}\label{7}
\begin{aligned}
n_{l,m}^\mathrm{max}&=\arg\min_{n_{1}}|\theta_{l,m}-\bar{\theta}_{n_{1}}|\overset{(a)}{=}\arg\min_{n_{1}}|\frac{f_{m}}{f_{c}}\psi_{l}-\bar{\theta}_{n_{1}}|\\
&\overset{(b)}{=}\arg\min_{n_{1}}|\frac{f_{m}}{f_{c}}\bar{\theta}_{n}-\bar{\theta}_{n_{1}}|,
\end{aligned}
\end{equation}
where $n_{1}\in\{1,2,\cdots,N\}$, $(a)$ comes from (\ref{2}) and $d=c/2f_{c}$, and $(b)$ comes from the assumption $\psi_{l}=\bar{\theta}_{n}$. Considering the definition of BSP in (\ref{6}) and (\ref{7}), we can conclude that the elements in BSP $\Xi_{n}$ are $(n_{l,m}^\mathrm{max},m),m=1,2,\cdots,M$. These elements are the element indexes of each column of matrix $\mathbf{Q}_{n}$ with the largest power. Denote $\psi_{l^{'}}=\bar{\theta}_{n^{'}}=\bar{\theta}_{n}+\frac{2b}{N}$ as a physical channel direction different from $\psi_{l}=\bar{\theta}_{n}$, where $b$ is a non-zero integer. Based on (\ref{7}), for the $M$-th subcarrier, we can obtain
\begin{equation}\label{9}
\begin{aligned}
n_{l^{'},m}^\mathrm{max}&=\arg\min_{n_{1}}|\frac{f_{M}}{f_{c}}(\bar{\theta}_{n}+\frac{2b}{N})-\bar{\theta}_{n_{1}}|\\
&=\arg\min_{n_{1}}|\frac{f_{M}}{f_{c}}\bar{\theta}_{n}-(\bar{\theta}_{n_{1}}-\frac{2bf_{M}}{Nf_{c}})|.
\end{aligned}
\end{equation}
Since $f_{M}>f_{c}$, we can simply get the element index $n_{l^{'},m}^\mathrm{max}$ with the largest power for the physical channel direction $\psi_{l^{'}}$ at the $M$-th subcarrier satisfies $|n_{l^{'},M}^\mathrm{max}-n_{l,M}^\mathrm{max}|\geq b$, based on (\ref{7}) and (\ref{9}). Thus, for a specific path component with the physical channel direction $\psi_{l}$, the power captured by the BSP $\Xi_{n}$ of $\mathbf{Q}_{n}$ satisfies
\begin{equation}\label{10}
\begin{aligned}
\|\mathbf{Q}_{n}&(\Xi_{n})\|_{2}^{2}\\
&=\sum_{m=1}^{M-1}\Gamma^{2}(\frac{f_{m}}{f_{c}}\bar{\theta}_{n}-n_{l,m}^\mathrm{max})+\Gamma^{2}(\frac{f_{M}}{f_{c}}\bar{\theta}_{n}-n_{l,M}^\mathrm{max})\\
&\overset{(a)}{>}\sum_{m=1}^{M-1}\Gamma^{2}(\frac{f_{m}}{f_{c}}\bar{\theta}_{n}-n_{l,m}^\mathrm{max})+\Gamma^{2}(\frac{f_{M}}{f_{c}}\bar{\theta}_{n}-n_{l^{'},M}^\mathrm{max})\\
&\geq\|\mathbf{Q}_{n}(\Xi_{n'})\|_{2}^{2},
\end{aligned}
\end{equation}
where $(a)$ comes from $|n_{l^{'},M}^\mathrm{max}-n_{l,M}^\mathrm{max}|\geq b$. (\ref{10}) implies that when the physical channel direction $\psi_{l}=\bar{\theta}_{n}$, the BSP $\Xi_{n}$ can capture the most power of $\mathbf{Q}_{n}$ compared with the BSP $\Xi_{n^{'}}$ of other physical channel directions $\bar{\theta}_{n'}$. This means the BSP $\Xi_{n}$ corresponds to a unique physical channel direction $\bar{\theta}_{n}$. Therefore, considering the BSP $\Xi_{n}$ is defined by the physical channel direction in (\ref{6}), we can conclude a one-to-one match between the physical channel direction $\bar{\theta}_{n}$ and the BSP $\Xi_{n}$.$\hfill\blacksquare$

\textbf{Lemma 1} indicates that the BSP $\Xi_{n}$ can be utilized to estimate the physical channel direction $\psi_{l}$, since it can be seen as a specific feature of the physical channel direction. Fig. 2 compares the BSP and the common support detection window utilized in existing schemes \cite{18,19} on physical channel direction estimation. We can see from Fig. 2 (a) that since the BSP $\Xi_{3}$ can exactly capture the channel power induced by channel path with physical channel direction $\bar{\theta}_{3}$, the correct physical channel direction $\psi_{1}=\bar{\theta}_{3}$ can be detected. However, when the common support detection window is exploited as shown in Fig. 2 (b), the common support detection window captures the most power at $\bar{\theta}_{4}$. This is because due to the beam split effect, the path with physical channel direction  $\psi_{1}=\bar{\theta}_{3}$ and $\psi_{2}=\bar{\theta}_{5}$ both generate channel power at $\bar{\theta}_{4}$. As a result, the common support detection window may cause estimation error. Then, the following \textbf{Lemma 2} provides some insights on how to determine the sparse support of the channel based on the BSP.

\begin{thm} \label{lemma2}
	Define $\bigcup_{b=-\Delta}^{\Delta}\Theta_{N}(\Xi_{n}+b)$ as the SDW which is generated by expanding the BSP $\Xi_{n}$, where $\Theta_{N}(\Xi_{n}+b)$ is defined as the set composed of elements as $(\Theta_{N}(a+b),e)$ when $(a,e)$ belongs to the BSP $\Xi_{n}$ with $\Theta_{N}(x)=\mod_{N}(x-1)+1$, and $\Delta$ is the size of the SDW. When the physical channel direction $\psi_{l}$ locates on the angle-domain samples $\bar{\theta}_{n}$ as $\psi_{l}=\bar{\theta}_{n}$, the ratio $\gamma$ between the power captured by the SDW of $\mathbf{Q}_{n}$ and the power of $\mathbf{Q}_{n}$ can be denoted as
	\begin{equation}\label{11}
	\gamma=\frac{1}{2N}\sum_{b=-\Delta}^{\Delta}\int_{-\frac{1}{N}}^{\frac{1}{N}}\Gamma^{2}\left(\Delta\theta-\frac{2b}{N}\right)\mathrm{d}\Delta\theta.
	\end{equation} 
\end{thm}
\emph{Proof:} Based on the definition of $\gamma$, we have
\begin{equation}\label{12}
\gamma=\frac{\sum_{b=-\Delta}^{\Delta}\|\mathbf{Q}_{n}(\Theta_{N}(\Xi_{n}+b))\|_{F}^{2}}{\|\mathbf{Q}_{n}\|_{F}^{2}}.
\end{equation}
Since $\|\mathbf{Q}_{n}\|_{F}^{2}=\sum_{m=1}^{M}\|\mathbf{q}_{l,m}\|_{F}^{2}=MN^{2}$ and the definition of the SDW is $\bigcup_{b=-\Delta}^{\Delta}\Theta_{N}(\Xi_{n}+b)$, we can obtain
\begin{equation}\label{12-2}
\gamma=\frac{1}{MN^{2}}\sum_{b=-\Delta}^{\Delta}\sum_{m=1}^{M}\Gamma^{2}(\theta_{l,m}-\bar{\theta}_{n_{l,m}^\mathrm{max}+b}).
\end{equation}
Then, considering $\bar{\theta}_{n_{l,m}^\mathrm{max}+b}=\bar{\theta}_{n_{l,m}^\mathrm{max}}+\frac{2b}{N}$ and denoting $\Delta\theta_{m}=\theta_{l,m}-\bar{\theta}_{n_{l,m}^\mathrm{max}}$, (\ref{12-2}) can be converted into
\begin{equation}\label{13}
\gamma=\frac{1}{MN^{2}}\sum_{b=-\Delta}^{\Delta}\sum_{m=1}^{M}\Gamma^{2}(\Delta\theta_{m}-\frac{2b}{N}).
\end{equation}
From (\ref{7}), we know that the range of $\Delta\theta_{m}$ is $\Delta\theta_{m}\in[-\frac{1}{N},\frac{1}{N}]$. Since the number of subcarriers $M$ is usually large (e.g., $M=512$), we assume $\Delta\theta_{m}$ distributes uniformly in its range, i.e., $\Delta\theta_{m}=-\frac{1}{N}+(m-1)\frac{2}{NM}$. Therefore, the summation on $m$ in (\ref{13}) can be rewritten by an integration form as
\begin{equation}\label{14}
\sum_{m=1}^{M}\Gamma^{2}(\Delta\theta-\frac{2b}{N})=\frac{MN}{2}\int_{-\frac{1}{N}}^{\frac{1}{N}}\Gamma^{2}(\Delta\theta-\frac{2b}{N})\mathrm{d}\Delta\theta.
\end{equation}
According to (\ref{13}) and (\ref{14}), (\ref{11}) can be proved.$\hfill\blacksquare$

\textbf{Lemma 2} indicates that the SDWs generated by expanding the BSP $\Xi_{n}$ can capture most power of $\mathbf{Q}_{n}$. For instance, when $f_{c}=100$ GHz, $B=15$ GHz, $N=256$, $M=512$, and $\psi_{l}=\bar{\theta}_{40}$, $\gamma=97.7\%$ power of $\mathbf{Q}_{40}$ can be captured by the SDW $\bigcup_{b=-4}^{4}\Theta_{N}(\Xi_{40}+b)$. In contrast, when the physical channel direction $\psi_{l}\neq\bar{\theta}_{n}$, e.g., $\psi_{l}=\bar{\theta}_{50}$, we can only capture $\gamma=0.6\%$ power of $\mathbf{Q}_{50}$ by utilizing the SDW $\bigcup_{b=-4}^{4}\Theta_{N}(\Xi_{40}+b)$. This observation means that after the physical channel direction $\psi_{l}$ is estimated by the BSP, the sparse channel supports at different subcarriers for the $l$-th path component can de directly obtained from the SDWs  determined by the BSP.

\subsection{Beam split pattern detection based channel estimation scheme}
Based on the BSP discussed above, we propose a beam split pattern detection (BSPD) based channel estimation scheme. The key idea is to estimate the physical channel direction $\psi_{l}$ by using the BSP at first, and then recover the sparse elements of the $l$-th path component by using the SDW generated by expanding the BSP. The above procedure will be carried out successively path by path until all path components are estimated.

\begin{algorithm}[htb]
	\caption{BSPD based channel estimation scheme}
	\label{alg:Framwork}
	\begin{algorithmic}[1]
		\REQUIRE ~~\\
		Observation matrix: $\mathbf{Y}$; Combining matrix: $\bar{\mathbf{A}}$\\
		Number of path components: $L$; SDWs size: $\Delta$\\
		\ENSURE ~~\\
		Estimated angle-domain channel $\hat{\mathbf{H}}=[\hat{\mathbf{h}}_{1},\hat{\mathbf{h}}_{2},\cdots,\hat{\mathbf{h}}_{M}]$\\	
		\STATE 
		$\mathbf{U}=[\mathbf{u}_{1},\mathbf{u}_{2},\cdots,\mathbf{u}_{M}]=\mathbf{Y}$\\
		\STATE
		$\Xi_{n}=\bigcup_{m=1}^{M}\left\{\left(\arg\min_{n_{1}}|\frac{f_{m}}{f_{c}}\bar{\theta}_{n}-\bar{\theta}_{n_{1}}|,m\right)\right\}$\\
		\FOR {$l\in\{1,2,\cdots,L\}$}
		\STATE
		$\mathbf{C}=\bar{\mathbf{A}}^{H}\mathbf{U}$\\
		\STATE
		$n_{l}^{*}=\arg\max_{n}\|\mathbf{C}(\Xi_{n})\|_{F}$
		\STATE
		$\Upsilon_{l}=\bigcup_{-\Delta}^{\Delta}\Theta_{N}(\Xi_{n_{l}^{*}}+\Delta)$
		\FOR {$m\in\{1,2,\cdots,M\}$}
		\STATE 
		$\Upsilon_{l,m}=\{i|(i,m)\in\Upsilon_{l}\}$\\
		\STATE
		$\hat{\mathbf{s}}_{l,m}=\mathbf{0}_{N\times1},\hat{\mathbf{s}}(\Upsilon_{l,m})=\bar{\mathbf{A}}^{\dagger}(:,\Upsilon_{l,m})\mathbf{u}_{m}$
		\STATE
		$\mathbf{u}_{m}=\mathbf{u}_{m}-\bar{\mathbf{A}}(:,\Upsilon_{l,m})\hat{\mathbf{s}}_{l,m}(\Upsilon_{l,m})$
		\ENDFOR
		\ENDFOR
		\FOR{$m\in\{1,2,\cdots,M\}$}
		\STATE
		$\Omega_{m}=\Upsilon_{1,m}\cup\Upsilon_{2,m}\cup\cdots\cup\Upsilon_{L,m}$
		\STATE
			$\hat{\mathbf{h}}_{m}=\mathbf{0}_{N\times1},\hat{\mathbf{h}}_{m}(\Omega_{m})=\bar{\mathbf{A}}^{\dagger}(:,\Omega_{m})\bar{\mathbf{y}}_{m}$
		\ENDFOR
		\RETURN $\hat{\mathbf{H}}=[\hat{\mathbf{h}}_{1},\hat{\mathbf{h}}_{2},\cdots,\hat{\mathbf{h}}_{M}]$
	\end{algorithmic}
\end{algorithm}

The pseudo-code of the proposed BSPD based channel estimation scheme is illustrated in \textbf{Algorithm 1}. Firstly, we initialize the residual matrix $\mathbf{U}\in\mathcal{C}^{PN_\mathrm{RF}\times M}$ as $\mathbf{U}=[\mathbf{u}_{1},\mathbf{u}_{2},\cdots,\mathbf{u}_{M}]=\mathbf{Y}$, where $\mathbf{u}_{m}$ denotes the residual for the $m$-th subcarrier. After that, we generate $N$ BSPs $\Xi_{n},n=1,2,\cdots,N$ according to (\ref{6}) in step $2$. Then, for the $l$-th path component, we estimate the physical channel direction $\psi_{l}$ based on \textbf{Lemma 1}. Specifically, inspired by the idea in OMP or SOMP based scheme, we first calculate the correlation matrix $\mathbf{C}$ as $\mathbf{C}=\bar{\mathbf{A}}^{H}\mathbf{U}$ in step $4$. In step $5$, we utilize BSPs to capture the power of the correlation matrix $\mathbf{C}$, and determine the index $n_{l}^{*}$ of the physical channel direction of the $l$-th path component as
\begin{equation}\label{16}
n_{l}^{*}=\arg\max_{n}\|\mathbf{C}(\Xi_{n})\|_{F}.
\end{equation}
Thanks to the one-to-one match between the physical channel direction and the BSP in $\textbf{Lemma 1}$, (\ref{16}) can guarantee the accuracy of the estimation on physical channel direction $\psi_{l}=\bar{\theta}_{n_{l}^{*}}$. After the physical channel direction $\psi_{l}$ is obtained, sparse channel supports at different subcarriers can be decided according to \textbf{Lemma 2}. In step $6$, we obtain the sparse channel supports $\Upsilon_{l}$ at different subcarriers for the $l$-th path component from the SDW generated by expanding the BSP, which is decided by the estimated physical channel direction $\bar{\theta}_{n_{l}^{*}}$ as $\Upsilon_{l}=\bigcup_{-\Delta}^{\Delta}\Theta_{N}(\Xi_{n_{l}^{*}}+\Delta)$. Then,  the influence of the $l$-th path component is removed to estimate remained path components. Specifically, we calculate the sparse channel support of the $l$-th path at the $m$-th subcarrier $\Upsilon_{l,m}$ in step $8$ as $\Upsilon_{l,m}=\{i|(i,m)\in\Upsilon_{l}\}$. After that, in step $9$, non-zero elements of the $l$-th path components at the $m$-th subcarrier $\hat{\mathbf{s}}_{l,m}$ are calculated according to the LS algorithm as
\begin{equation}\label{17}
\hat{\mathbf{s}}_{l,m}=\mathbf{0}_{N\times1},\quad\hat{\mathbf{s}}(\Upsilon_{l,m})=\bar{\mathbf{A}}^{\dagger}(:,\Upsilon_{l,m})\mathbf{u}_{m}.
\end{equation}
In step $10$, we can remove the influence of the $l$-th path and update the residual matrix as
\begin{equation}\label{17-1}
\mathbf{u}_{m}=\mathbf{u}_{m}-\bar{\mathbf{A}}(:,\Upsilon_{l,m})\hat{\mathbf{s}}_{l,m}(\Upsilon_{l,m}).
\end{equation}

The procedure above is carried out $L$ times until the sparse channel supports of all path components are estimated, where the number of path components $L$ can be obtained from channel measurement in advance \cite{28}. Finally, the angle-domain channel $\bar{\mathbf{h}}_{m},m=1,2,\cdots,M$ is estimated based on these sparse channel supports. In specific, we calculate the sparse channel support of the $m$-th subcarrier as $\Omega_{m}=\Upsilon_{1,m}\cup\Upsilon_{2,m}\cup\cdots\cup\Upsilon_{L,m}$ in step $14$. Then, we could obtain the estimated sparse angle-domain channel as
\begin{equation}\label{17-2} \hat{\mathbf{h}}_{m}=\mathbf{0}_{N\times1},\hat{\mathbf{h}}_{m}(\Omega_{m})=\hat{\mathbf{A}}^{\dagger}\bar{\mathbf{y}}_{m},
\end{equation}
where $\bar{\mathbf{y}}_{m}$ represents the overall received pilots at the $m$-th subcarrier.

Notice that although the proposed scheme is inspired by the SOMP algorithm in \cite{18,31}, the proposed scheme has a major difference from the SOMP based scheme on how to detect the sparse channel supports. Specifically, in SOMP based scheme, the sparse supports are supposed to satisfy the common sparse support assumption. Therefore, the sparse supports are detected column by column where each column corresponds to a physical channel direction. On the contrary, the proposed scheme utilizes a two-step procedure to obtain sparse channel supports. Firstly, the  physical channel directions are detected by using the defined BSP to capture the power of the channel. The elements in the BSP are determined by the beam split effect and not locate on a common position, which is quite different from the SOMP based scheme. Secondly, after the physical channel directions are obtained, the sparse supports are generated by expanding the BSPs which are determined by the obtained physical channel directions. Due to the beam split effect in the wideband THz massive MIMO channel, the ideal common sparse support assumption is not reasonable. Thus, the SOMP based scheme will face performance loss. While, since the proposed scheme makes use of the frequency-dependent sparse channel supports implied by the beam split effect, it can correctly detect the physical channel directions and corresponding sparse channel supports. Hence, the proposed scheme can achieve a better channel estimation accuracy.

\subsection{Performance analysis}\label{PA}
In this subsection, we will analyze the estimation accuracy on physical channel directions $\psi_{l}$ of our proposed BSPD based channel estimation scheme. We prove that the physical channel directions $\psi_{l}$ with $l=1,2,\cdots,L$ can be estimated precisely with a certain probability. Note that in the following analysis, we assume the physical channel direction for the $l$-th path component locates on the angle-domain samples satisfying $\psi_{l}=\bar{\theta}_{n_{l}}$, which will only lead to a negligible physical channel direction estimation error when the number of antennas is huge in THz massive MIMO systems.

Firstly, we rewrite the angle-domain channel $\mathbf{H}$ in (\ref{15}) to decouple different path components in $\mathbf{H}$. Specifically, we define $\Upsilon_{n}$ as the set containing all the index $i$ in the $n$-th BSP $\Xi_{n}$ with $\Upsilon_{n}=\{i|(i,m)\in\Xi_{n},m=1,2,\cdots,M\}$. For the $l$-th path component, since the rows indexed by $\Upsilon_{n_{l}}$ contains all the elements in BSP $\Xi_{n_{l}}$, most of its power can be captured by $\mathbf{Q}_{n_{l}}(\Upsilon_{n_{l}},:)$. Based on this property, we rewrite the angle-domain channel $\mathbf{H}$ as $\mathbf{H}=\mathbf{V}\mathbf{B}$. The matrix $\mathbf{B}\in\mathcal{C}^{\sum_{n=1}^{N}|\Upsilon_{n}|\times M}$ is defined as $\mathbf{B}=[\mathbf{B}_{1}^{H},\mathbf{B}_{2}^{H},\cdots,\mathbf{B}_{N}^{H}]$, where
\begin{equation}\label{18}
\mathbf{B}_{n}=\left\{
\begin{aligned}
&\mathbf{H}(\Upsilon_{n},:),\quad n\in\{n_{1},n_{2},\cdots,n_{L}\},\\
&\mathbf{0}_{|\Upsilon_{n}|\times M},\quad n\notin\{n_{1},n_{2},\cdots,n_{L}\},
\end{aligned}
\right.
\end{equation}
with $\mathbf{H}(\Upsilon_{n_{l}},:)=g_{l}\mathbf{Q}_{n_{l}}(\Upsilon_{n_{l}},:)$. Correspondingly, $\mathbf{V}\in\mathcal{C}^{N\times\sum_{n=1}^{N}|\Upsilon_{n}|}$ becomes a transformation matrix with $\mathbf{V}=[\mathbf{V}_{1},\mathbf{V}_{2},\cdots,\mathbf{V}_{N}]$, where the $i$-th column of   $\mathbf{V}_{n}\in\mathcal{C}^{N\times|\Upsilon_{n}|}$ only has one nonzero element at the index $\Upsilon_{n}(i)$ with $\Upsilon_{n}(i)$ representing the $i$-th element in the set $\Upsilon_{n}$. This transformation $\mathbf{H}=\mathbf{V}\mathbf{B}$ can convert the angle-domain channel $\mathbf{H}$ into a block-wise form, where each path component is corresponding to a specific block in $\mathbf{B}$.

With the help of the transformation $\mathbf{H}=\mathbf{V}\mathbf{B}$, we can rewrite (\ref{15}) as
\begin{equation}\label{19}
\mathbf{Y}=\left[\bar{\mathbf{A}}(:,\Upsilon_{1}),\bar{\mathbf{A}}(:,\Upsilon_{2}),\cdots,\bar{\mathbf{A}}(:,\Upsilon_{N})\right]\mathbf{B}+\mathbf{N}.
\end{equation}
Then, to estimate the physical channel directions $\psi_{l}$, the key correlation matrix $\mathbf{C}$ is calculated as described in step $4$ of \textbf{Algorithm 1}. Considering the transformation in (\ref{18}), the correlation matrix $\mathbf{C}$ can be denoted as
\begin{equation}\label{20}
\mathbf{C}=\bar{\mathbf{A}}^{H}\mathbf{Y}=\sum_{i=1}^{N}\bar{\mathbf{A}}^{H}\bar{\mathbf{A}}(:,\Upsilon_{i})\mathbf{B}_{i}+\bar{\mathbf{A}}^{H}\mathbf{N}.
\end{equation}
Therefore, the power captured by the BSP $\Xi_{n}$, which is utilized to estimate the physical channel direction $\psi_{l}$ as shown in step $5$ of \textbf{Algorithm 1}, can be represented as
\begin{equation}\label{21}
\begin{aligned}
\|\mathbf{C}(\Xi_{n})\|_{F}=\sum_{m=1}^{M}\bigg|\sum_{i=1}^{N}\bar{\mathbf{A}}^{H}&(:,\Xi_{n}(m))\bar{\mathbf{A}}(:,\Upsilon_{i})\mathbf{B}_{i}(:,m)\\
&+\bar{\mathbf{A}}^{H}(:,\Xi_{n}(m))\mathbf{N}(:,m)\bigg|,
\end{aligned}
\end{equation}
where $\Xi_{n}(m)$ denotes the index $a$ that satisfies $(a,m)\in\Xi_{n}$. To illustrate the analysis clearly, we define an auxiliary parameter $\mu$ to represent the sub-coherence of the matrix $\bar{\mathbf{A}}$ as
\begin{equation}\label{22}
\mu\triangleq\max_{i,j\in{1,2,\cdots,N},i\neq j}|\bar{\mathbf{A}}^{H}(:,i)\bar{\mathbf{A}}(:,j)|.
\end{equation}
Note that the sub-coherence of the observation matrix $\bar{\mathbf{A}}$ defined in (\ref{22}) is widely utilized in the performance analysis of compressive sensing based algorithms \cite{32}.

Based on the definitions above, we prove the following \textbf{Lemma 3}, which provides a lower bound of the correct probability of the physical channel direction estimation for the proposed BSPD based scheme. Specifically, for a certain channel and a noise level, the physical channel direction $\psi_{l}$ can be accurately estimated with a probability larger than a certain probability. The specific description and proof of \textbf{Lemma 3} are illustrated as follows.

\begin{thm} \label{lemma3}
	For the $l$-th path component, we assume the physical channel direction $\psi_{l}$ locates on the angle-domain samples as $\psi_{l}=\bar{\theta}_{n_{l}}$. When
	\begin{equation}\label{23}
	\begin{aligned}
	\sum_{m=1}^{M}&\mathbf{B}_{n_{l}}(\chi(\Xi_{n_{l}}(m)),m)-\mu\sqrt{|\Upsilon_{n_{l}}|}\|\mathbf{B}_{n_{l}}\|_{F}\\
	&-2\mu\sum_{n_{i}\in\mathcal{L}\backslash n_{l}}\sqrt{|\Upsilon_{n_{i}}|}\|\mathbf{B}_{n_{i}}\|_{F}
	\geq 2\sqrt{M\sigma^{2}\alpha}
	\end{aligned}
	\end{equation} 
	holds with $\chi(\cdot)$ denoting the index transformation as $\chi(\Xi_{n_{l}}(m))=\Xi_{n_{l}}(m)-\min\Upsilon_{n_{l}}+1$ and $\mathcal{L}=\{n_{1},n_{2},\cdots,n_{L}\}$, the proposed BSPD based channel estimation scheme can accurately estimate $\psi_{l}$ with a probability exceeding
	\begin{equation}\label{24}
	\left(1-0.8\alpha^{-\frac{1}{2}}e^{-\frac{\alpha}{2}}\right)^{M},
	\end{equation}
	with $\alpha$ is a constant.
\end{thm}
\emph{Proof:} See Appendix A.$\hfill\blacksquare$

\textbf{Lemma 3} indicates that for a certain channel and a noise level, the correct probability of the physical channel direction estimation can be lower-bounded by (\ref{24}). We can observe from (\ref{24}) that when the correct probability in (\ref{24}) locates in the feasible domain as $0\leq\left(1-0.8\alpha^{-\frac{1}{2}}e^{-\frac{\alpha}{2}}\right)^{M}\leq 1$, the correct probability will monotonically increase from $0$ to $1$ rapidly when $\alpha$ grows up. Therefore, considering that the allowed $\alpha$ is large given $\mu$ and $\|\mathbf{B}_{n}\|_{F}$ when the noise power $\sigma^{2}$ is relatively small, the correct probability of the physical channel direction estimation will approach $1$ with a low noise level. Therefore, we can conclude that the proposed BSPD based channel estimation scheme is able to accurately estimate physical channel directions when the noise power is relatively small. This conclusion will be verified by simulation results in Section \ref{Sim}.

\subsection{Complexity analysis}
In this subsection, we will provide the complexity analysis of the proposed BSPD based channel estimation scheme based on the number of complex multiplications. The complexity of the OMP and SOMP based schemes \cite{18,19} are also shown for comparison.

We can observe from \textbf{Algorithm 1} that the complexity of the proposed BSPD based scheme is mainly determined by steps $4$, $5$, $9$, $10$, and $15$. Specifically, in step $4$, the correlation matrix $\mathbf{C}$ is calculated by $\mathbf{C}=\bar{\mathbf{A}}^{H}\mathbf{U}$. Since the dimension of $\bar{\mathbf{A}}$ and $\mathbf{U}$ are $\bar{\mathbf{A}}\in\mathcal{C}^{N_\mathrm{RF}P\times N}$ and
$\mathbf{U}\in\mathcal{C}^{PN_\mathrm{RF}\times M}$, we know that the complexity of step $4$ is $\mathcal{O}(N_\mathrm{RF}PNM)$. Then, we compute the norm of $\mathbf{C}(\Xi_{n})$ of size $M\times 1$ for $N$ times in step $5$. Therefore, the complexity of step $5$ should be $\mathcal{O}(MN)$. In step $9$, the non-zero elements at each subcarrier $m$ are calculated as (\ref{17}), where the pseudo-inverse of $\bar{\mathbf{A}}(:,\Upsilon_{l,m})\in\mathcal{C}^{N_\mathrm{RF}P\times(2\Delta+1)}$ is generated together with the multiplication between  $\bar{\mathbf{A}}^{\dagger}(:,\Upsilon_{l,m})$ and $\mathbf{u}_{m}\in\mathcal{C}^{N_\mathrm{RF}P\times 1}$. Consequently, step $9$ requires the complexity $\mathcal{O}(MN_\mathrm{RF}P\Delta^{2})$. Similarly, the complexity of step $10$, which contains a multiplication between $\bar{\mathbf{A}}^{\dagger}(:,\Upsilon_{l,m})$ and $\hat{\mathbf{s}}_{l,m}\in\mathcal{C}^{N_\mathrm{RF}P\times 1}$, is $\mathcal{O}(MN_\mathrm{RF}P\Delta)$. In addition, the sparse angle-domain channel at each subcarrier $m$ is recovered in step $15$ with a multiplication between $\bar{\mathbf{A}}^{\dagger}(:,\Omega_{m})\in\mathcal{C}^{N_\mathrm{RF}P\times(L(2\Delta+1))}$ and $\mathbf{y}_{m}\in\mathcal{C}^{N_\mathrm{RF}P\times1}$. Therefore, the step $15$ involves the complexity of $\mathcal{O}(MN_\mathrm{RF}PL^{2}\Delta^{2})$. Finally, considering that steps $4$, $5$, $9$, and $10$ are carried out for $L$ times, the total complexity of the proposed BSPD based scheme can be represented as
\begin{equation}\label{Com}
\begin{aligned}
\mathcal{O}&(MNN_\mathrm{RF}PL)+\mathcal{O}(MNL)\\
&+\mathcal{O}(MN_\mathrm{RF}PL\Delta^{2})+\mathcal{O}(MN_\mathrm{RF}PL^{2}\Delta^{2}).
\end{aligned}
\end{equation}

For comparison, both the OMP and SOMP based channel estimation schemes require the complexity of $\mathcal{O}(MNN_\mathrm{RF}PL\Delta)+\mathcal{O}(MN_\mathrm{RF}PL^{3}\Delta^{3})$ \cite{18,19}. Generally, due to the sparsity of the THz channel, the size of the SDW is relatively small, e.g., $\Delta=4\ll N=256$. Therefore, we can conclude that the proposed BSPD based scheme enjoys a lower complexity than the existing OMP and the SOMP based schemes \cite{18,19}.

\section{Simulation Results}\label{Sim}
In this section, we provide simulation results for the proposed BSPD based channel estimation scheme. We consider a multi-user THz wideband MIMO-OFDM system. The system parameters are set as: $N=256$, $K=8$, $N_\mathrm{RF}=8$, $f_{c}=100$ GHz, $B=15$ GHz and $M=512$. Note that $B=15$ GHz is a reasonable setting for THz systems to reach target transmission rate as $1$ Tb/s in future 6G networks \cite{3}. The multipath THz channel is generated with the following parameters: $L=3$, $g_{l}\sim\mathcal{CN}(0,1)$, $\bar{\psi}_{l}\sim\mathcal{U}(-\frac{\pi}{2},\frac{\pi}{2})$ and $\max_{l}\tau_{l}=20$ ns. Finally, the signal-to-noise ratio (SNR) is defined as $\frac{1}{\sigma^{2}}$. 

\begin{figure}
	\centering
	\includegraphics[width=0.45\textwidth]{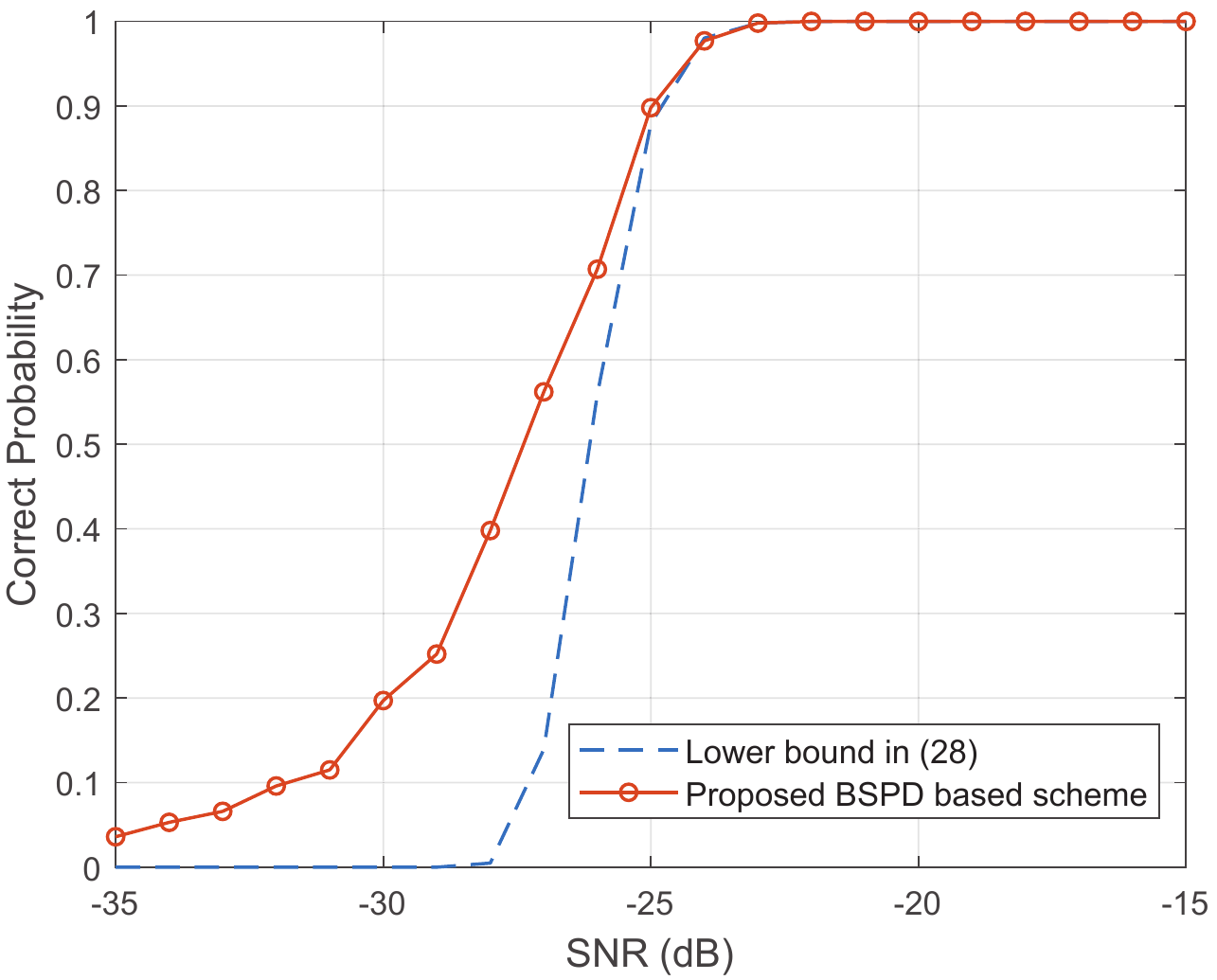}
	\caption{The correct probability of the physical channel direction estimation of the proposed BSPD based channel estimation scheme.}
\end{figure}

Fig. 3 provides the physical channel direction estimation accuracy of the proposed BSPD based scheme. We set the physical channel directions of the channel path components as $\psi_{1}=\bar{\theta}_{193}=0.5039, \psi_{2}=\bar{\theta}_{193}=-0.8711, \psi_{3}=\bar{\theta}_{193}=-0.3008$. The probability that the proposed BSPD based scheme could correctly estimate the physical channel direction $\psi_{1}$ is shown in Fig. 3. Besides, the lower bound of the correct probability of the physical channel direction estimation, which is proved by \textbf{Lemma 3}, is also illustrated for comparison. From Fig. 3, we can observe that when the SNR is larger than $-25$ dB, the proposed BSPD based scheme has the ability to obtain the correct physical channel direction with a probability of $1$. This indicates that the proposed scheme can estimate the physical channel direction accurately even in low SNR regions. Moreover, we can observe that the correct probability of the physical channel direction estimation achieved by the proposed scheme is tightly lower-bounded by the bound proved in \textbf{Lemma 3}, which is consistent with our analysis in Subsection \ref{PA}.

\begin{figure}
	\centering
	\includegraphics[width=0.45\textwidth]{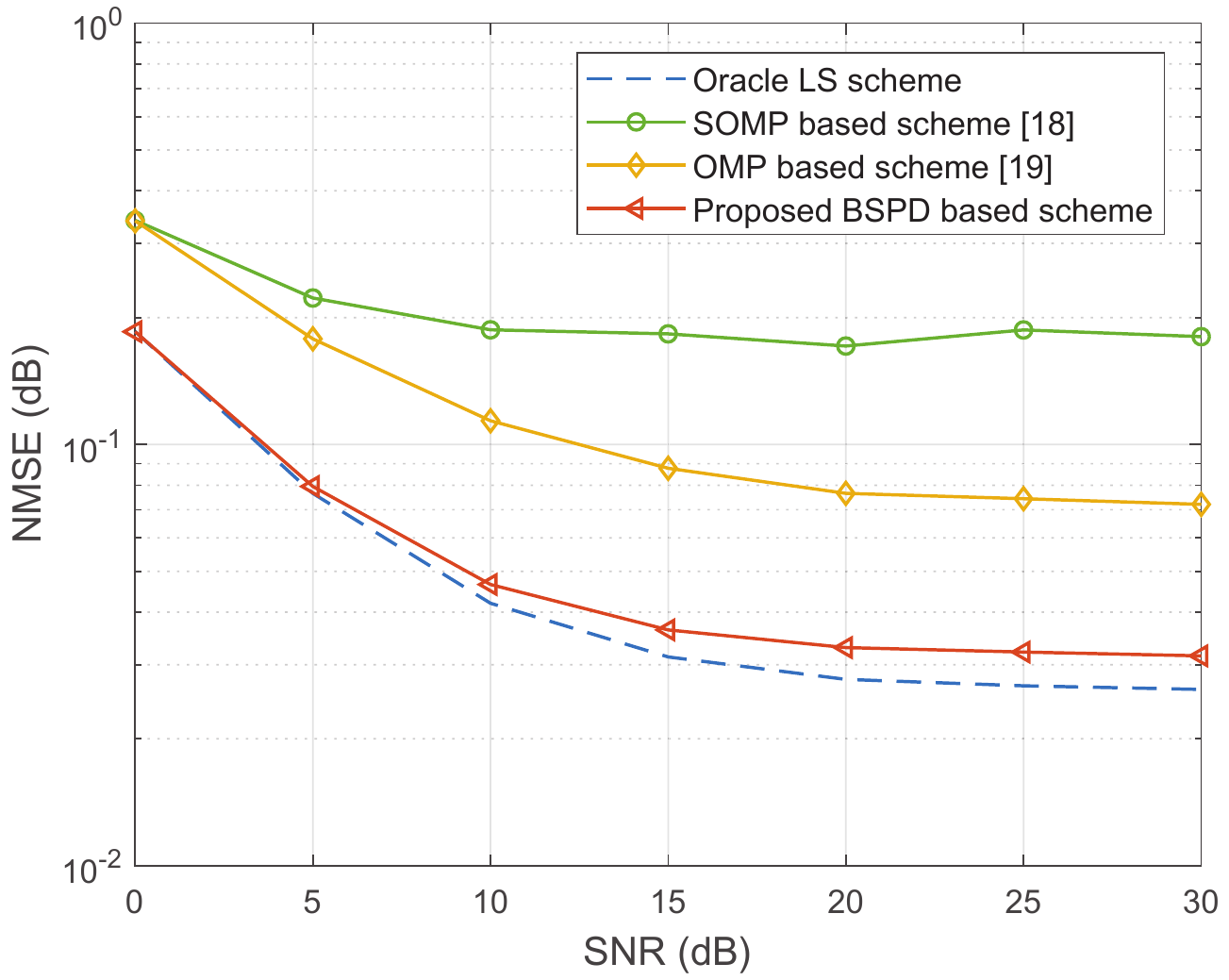}
	\caption{NMSE performance comparison against SNR.}
\end{figure}

Fig. 4 illustrates the normalized mean square error (NMSE) performance against SNR of the proposed BSPD based scheme and existing schemes, including SOMP based scheme \cite{18} and OMP based scheme \cite{19}. For the OMP based scheme, we carry out the OMP algorithm once every $16$ subcarrier.Then, the sparse channel supports of these $16$ subcarriers are obtained by the OMP algorithm based on the common support assumption. For all considered schemes, $P=10$ time slots per user are utilized for pilot transmission. The size of SDWs is set as $\Delta=4$ for the proposed BSPD based scheme. For a fair comparison, we assume the sparsity level in OMP based scheme and SOMP based scheme is $L(2\Delta+2)=27$. The oracle LS scheme is also considered as a benchmark for comparison, where the sparse channel supports of the wideband channel $\bar{\mathbf{h}}_{m}$ are assumed to be known perfectly. We can observe from Fig. 4 that the proposed BSPD scheme outperforms existing schemes \cite{18,19} in all SNR regions. This is because the BSPD based scheme exploits the specific sparse property of the wideband channel with the beam slit effect. Moreover, our proposed BSPD scheme can approach the NMSE performance of the ideal oracle LS scheme.

\begin{figure}
	\centering
	\includegraphics[width=0.45\textwidth]{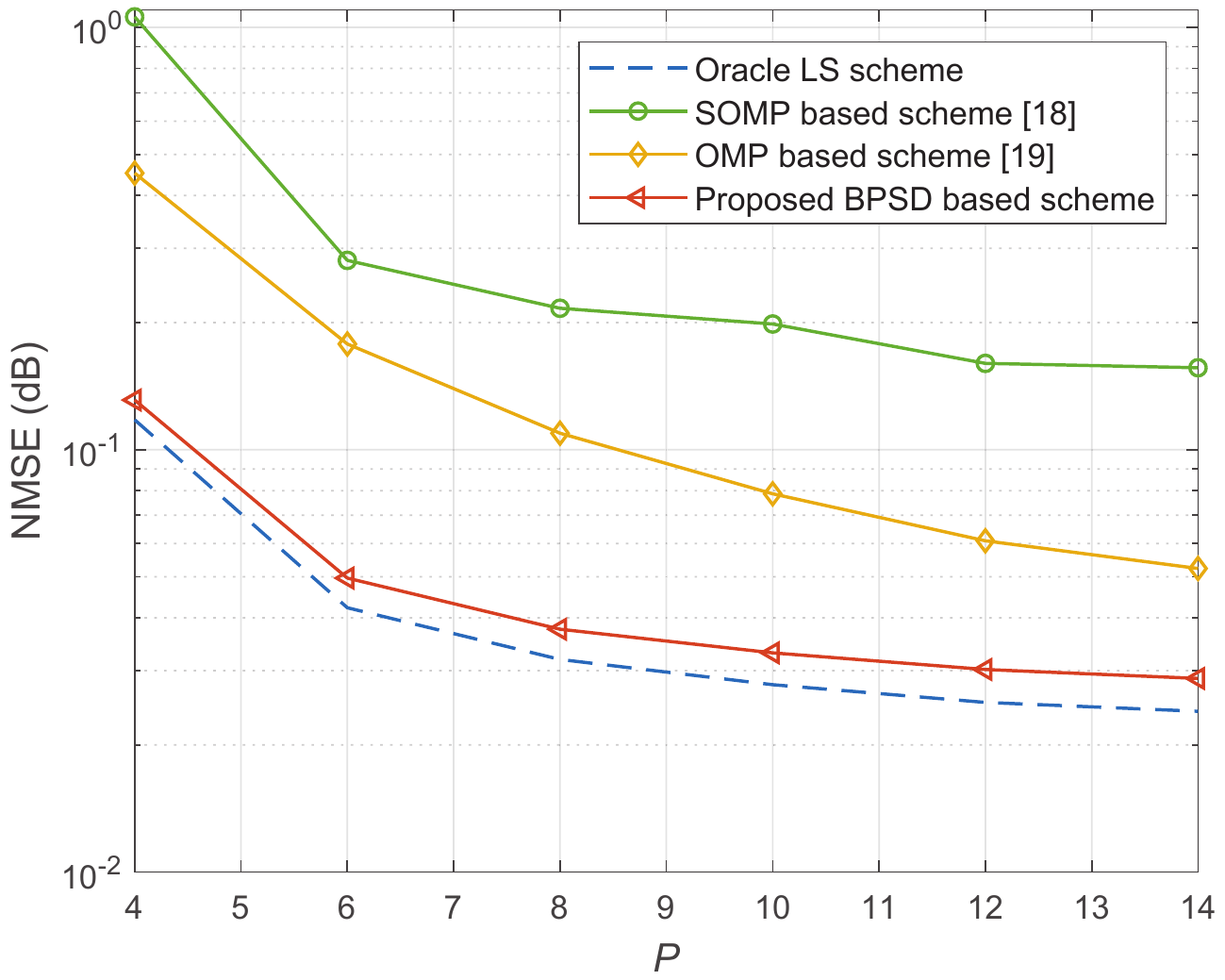}
	\caption{NMSE performance comparison against the pilot length $P$.}
\end{figure}
Fig. 5 provides the NMSE performance against pilot length $P$, where SNR is set as $20$ dB. Other parameters are the same as those in Fig. 4. We can observe from Fig. 5 that the NMSE achieved by all schemes decrease as the pilot sequence becomes longer. In all considered length of pilots $P$, the proposed BSPD based scheme can achieve better NMSE performance than existing schemes, and it can approach the NMSE performance of the ideal oracle LS scheme. Particularly, in the case with a short length of pilots (e.g., $P=4$ and $P=8$), the performance gap between the proposed BSPD based scheme and existing schemes is quite large. This indicates that the BSPD based scheme can efficiently reduce the pilot overhead for channel estimation.

Fig. 6 shows the NMSE performance against the bandwidth $B$. The range of the bandwidth is from $1$ GHz to $15$ GHz, and other parameters are set as SNR $=20$ dB and $P=10$. We can observe from Fig. 6 that when the bandwidth is small, e.g., $1$ GHz, both the SOMP based scheme and the proposed scheme can achieve the near-optimal NMSE performance. However, when the bandwidth becomes larger, the NMSE performance of the SOMP and OMP based scheme gradually degrade. This is because the assumption of a common sparse channel support utilized in the SOMP and OMP based schemes cannot deal with the frequency-dependent sparse channel supports caused by the beam split effect. In contrast, the proposed BSPD based scheme is robust to the bandwidth $B$, and can achieve the near-optimal NMSE performance with different bandwidths.

\begin{figure}
	\centering
	\includegraphics[width=0.45\textwidth]{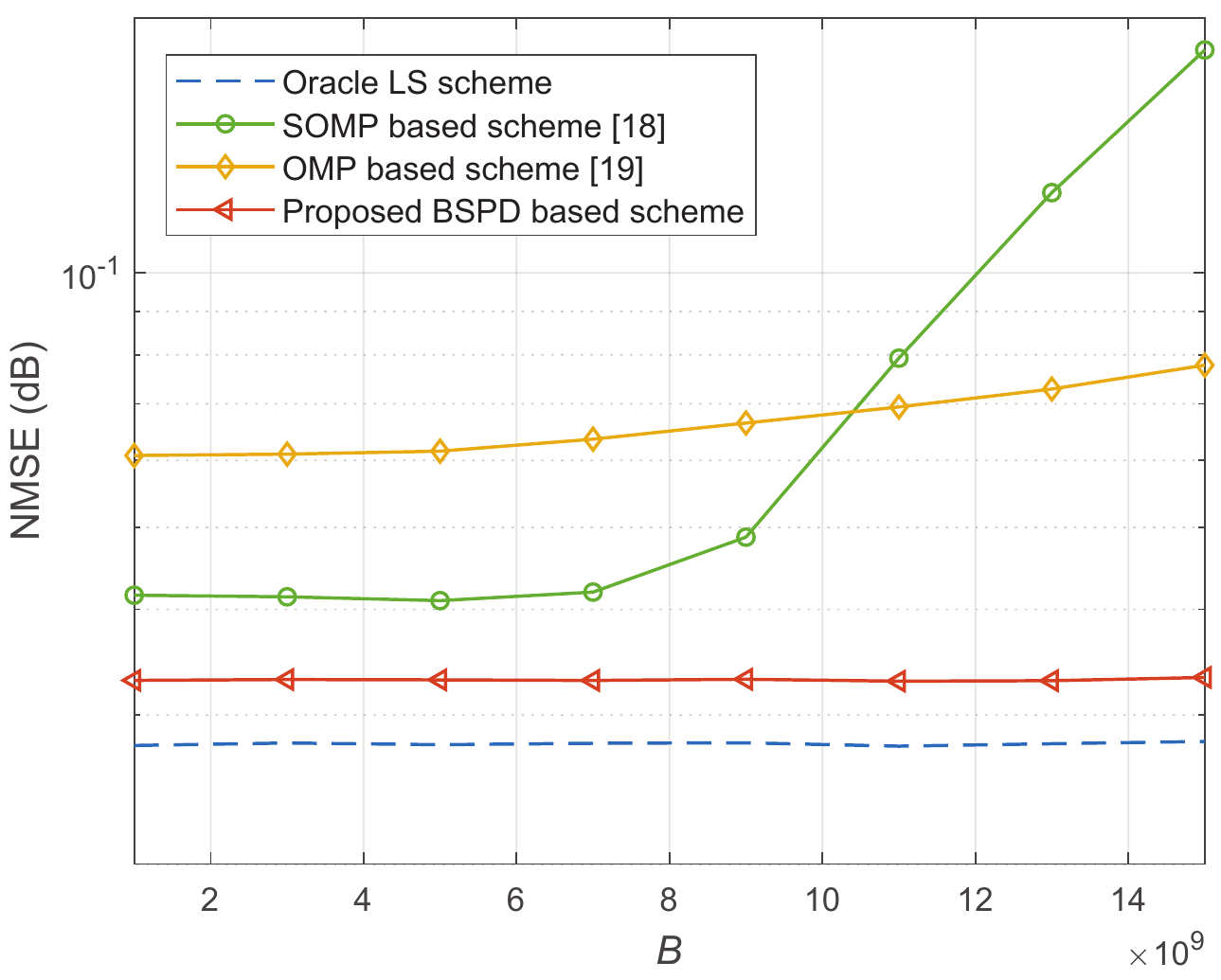}
	\caption{NMSE performance comparison against the bandwidth $B$.}
\end{figure}

\section{CONCLUSION}\label{Con}
In this paper, we investigated the channel estimation problem in wideband THz systems, where the beam split effect was considered. We proposed an efficient BSPD based wideband channel estimation scheme. Firstly, we proved the one-to-one match between the physical channel direction and the BSP, and the SDW generated by expanding the BSP corresponds to the sparse supports of the channel path component. Based on this proof, we propose to estimate the physical channel directions by using the BSPs, and then recover the sparse channel supports by exploiting the SDWs generated by expanding the BSPs. Simulation results show that the proposed scheme can achieve better NMSE performance than existing schemes.

\section*{ACKNOWLEDGEMENT}
\label{ACKNOWLEDGEMENT}
This work was supported in part by the National Key Research and Development Program of China (Grant No. 2020YFB1805005), in part by the National Natural Science Foundation of China (Grant No. 62031019), and in part by the European Commission through the H2020-MSCA-ITN META WIRELESS Research Project under Grant 956256.

\section*{Appendix A. Proof of Lemma 3}
\emph{Proof:} For the $l$-th path component, the physical channel direction $\psi_{l}$ can be accurately estimated if 
\begin{equation}\label{25}
\|\mathbf{C}(\Xi_{n_{l}})\|_{F}\geq\max_{n\notin\mathcal{L}}\|\mathbf{C}(\Xi_{n})\|_{F}.
\end{equation}
We consider a specific case $\mathcal{D}$ that the noise term in (\ref{21}) is bounded by a constant $\gamma$ as
\begin{equation}\label{26}
\begin{aligned}
&\mathcal{D}=\\
&\left\{|\bar{\mathbf{A}}^{H}(:,\Xi_{n}(m))\mathbf{N}(:,m)|^{2}\leq\gamma,m=1,2,\cdots,M\right\},
\end{aligned}
\end{equation}
where $\gamma$ is defined as $\gamma=\sigma^{2}\alpha$. Based on \textbf{Lemma 4} proved in Appendix B, the case $\mathcal{D}$ will occur with a probability exceeding (\ref{24}).

In this case, we can bound the right side of (\ref{25}) according to (\ref{21}) as
\begin{equation}\label{27}
\begin{aligned}
\max_{n\notin\mathcal{L}}|\mathbf{C}(\Xi_{n})\|_{F}\leq&\max_{n\notin\mathcal{L}}\sum_{m=1}^{M}|\bar{\mathbf{A}}^{H}(:,\Xi_{n}(m))\mathbf{N}(:,m)|\\
&+\max_{n\notin\mathcal{L}}\sum_{m=1}^{M}\sum_{n_{i}\in\mathcal{L}}\|\mathbf{B}_{n_{i}}(:,m)\|_{F}\\
&\times\|\bar{\mathbf{A}}^{H}(:,\Xi_{n}(m))\bar{\mathbf{A}}(:,\Upsilon_{n_{i}})\|_{F}.
\end{aligned}
\end{equation}
Considering the noise matrix $\mathbf{N}$ is composed of $M$ vectors with Gaussian distribution and case $\mathcal{D}$ happens, we can obtain
\begin{equation}\label{28}
|\bar{\mathbf{A}}^{H}(:,\Xi_{n}(m))\mathbf{N}(:,m)|^{2}\leq\gamma=\sigma^{2}\alpha.
\end{equation}
In addition, due to the definition of $\mu$ in (\ref{22}), we have $\|\bar{\mathbf{A}}^{H}(:,\Xi_{n}(m))\bar{\mathbf{A}}(:,\Upsilon_{n_{i}})\|_{F}\leq\sqrt{|\Upsilon_{n_{i}}|}\mu$. Therefore, considering $\sum_{m=1}^{M}\|\mathbf{B}_{n_{i}}(:,m)\|_{2}=\|\mathbf{B}_{n_{i}}\|_{F}$, we can obtain the upper bound of the right side of (\ref{25}) as
\begin{equation}\label{29}
\max_{n\notin\mathcal{L}}\|\mathbf{C}(\Xi_{n})\|_{F}\leq \sqrt{M\sigma^{2}\alpha}+\mu\sum_{n_{i}\in\mathcal{L}}\sqrt{|\Upsilon_{n_{i}}|}\|\mathbf{B}_{n_{i}}\|_{F}.
\end{equation}

On the other hand, we can obtain the lower bound of the left side of (\ref{25}) as
\begin{equation}\label{30}
\begin{aligned}
&\|\mathbf{C}(\Xi_{n_{l}})\|_{F}\\
&\geq\sum_{m=1}^{M}\bigg(|\bar{\mathbf{A}}^{H}(:,\Xi_{n_{l}}(m))\bar{\mathbf{A}}(:,\Upsilon_{n_{l}})\mathbf{B}_{n_{l}}(:,m)|\\
&-|\bar{\mathbf{A}}^{H}(:,\Xi_{n_{l}}(m))\mathbf{N}(:,m)|\\
&-\sum_{n_{i}\in\mathcal{L}\backslash n_{l}}|\bar{\mathbf{A}}^{H}(:,\Xi_{n_{l}}(m))\bar{\mathbf{A}}(:,\Upsilon_{n_{i}})\mathbf{B}_{n_{i}}(:,m)|\bigg).
\end{aligned}
\end{equation}
Based on the definition of the analog combining matrix $\bar{\mathbf{A}}$, we know that $|\bar{\mathbf{A}}^{H}(:,i)\bar{\mathbf{A}}(:,i)|=1$ and $|\bar{\mathbf{A}}^{H}(:,i)\bar{\mathbf{A}}(:,j)|\geq 0$ for $i\neq j$. Hence, the first term in the right side of (\ref{30}) can be bounded as
\begin{equation}\label{31}
\begin{aligned}
\sum_{m=1}^{M}|\bar{\mathbf{A}}^{H}(:,\Xi_{n_{l}}(m))&\bar{\mathbf{A}}(:,\Upsilon_{n_{l}})\mathbf{B}_{n_{l}}(:,m)|\\
&\geq\sum_{m=1}^{M}|\mathbf{B}_{n_{l}}(\chi(\Xi_{n_{l}}(m)),m)|,
\end{aligned}
\end{equation}
where $\chi(\cdot)$ denotes the index transformation as $\chi(\Xi_{n_{l}}(m))=\Xi_{n_{l}}(m)-\min\Upsilon_{n_{l}}+1$. Similar to (\ref{28}), we can bound the second term in the right side of (\ref{30}) as $-|\bar{\mathbf{A}}^{H}(:,\Xi_{n_{l}}(m))\mathbf{N}(:,m)|\geq-\sqrt{\sigma^{2}\alpha}$. Then, according to the operation on the second term in the right side of (\ref{27}), the lower bound of the third term in the right side of (\ref{30}) can be represented as
\begin{equation}\label{32}
\begin{aligned}
-\sum_{m=1}^{M}\sum_{n_{i}\in\mathcal{L}\backslash n_{l}}|\bar{\mathbf{A}}^{H}&(:,\Xi_{n_{l}}(m))\bar{\mathbf{A}}(:,\Upsilon_{n_{i}})\mathbf{B}_{n_{i}}(:,m)|\\
&\geq-\mu\sum_{n_{i}\in\mathcal{L}\backslash n_{l}}\sqrt{|\Upsilon_{n_{i}}|}\|\mathbf{B}_{n_{i}}\|_{F}.
\end{aligned}
\end{equation}
Combining the above analysis in (\ref{31}) and (\ref{32}), we can conclude that
\begin{equation}\label{33}
\begin{aligned}
\|\mathbf{C}(\Xi_{n_{l}})\|_{F}\geq&\sum_{m=1}^{M}\mathbf{B}_{n_{l}}(\chi(\Xi_{n_{l}}(m)),m)\\
&-\sqrt{M\sigma^{2}\alpha}-\mu\sum_{n_{i}\in\mathcal{L}\backslash n_{l}}\sqrt{|\Upsilon_{n_{i}}|}\|\mathbf{B}_{n_{i}}\|_{F}.
\end{aligned}
\end{equation}
Therefore, based on the bounds in (\ref{29}) and (\ref{33}), we can conclude that when 
\begin{equation}\label{34}
\begin{aligned}
\sum_{m=1}^{M}&\mathbf{B}_{n_{l}}(\chi(\Xi_{n_{l}}(m)),m)-\mu\sqrt{|\Upsilon_{n_{l}}|}\|\mathbf{B}_{n_{l}}\|_{F}\\
&-2\mu\sum_{n_{i}\in\mathcal{L}\backslash n_{l}}\sqrt{|\Upsilon_{n_{i}}|}\|\mathbf{B}_{n_{i}}\|_{F}\geq \sqrt{2M\sigma^{2}\alpha},
\end{aligned}
\end{equation}
(\ref{25}) is guaranteed under the case $\mathcal{D}$, and consequently the physical channel direction $\psi_{l}$ can be accurately estimated with a probability exceeding (\ref{24}). Thus, the proof is completed.$\hfill\blacksquare$

\section*{Appendix B. Lemma 4}
\begin{thm} \label{lemma4}
	Assuming each column of the noise matrix $\mathbf{N}$ in (\ref{15}) is a Gaussian vector satisfying $\mathbf{N}(:,m)\sim\mathcal{CN}(\mathbf{0}_{QN_\mathrm{RF}},\sigma^{2}\mathbf{I}_{QN_\mathrm{RF}})$, we have the probability that the case $\mathcal{D}$ in (\ref{26}) happens satisfies
	\begin{equation}\label{App1}
	\mathrm{Pr}\left\{\mathcal{D}\right\}\geq \left(1-0.8\alpha^{-\frac{1}{2}}e^{-\frac{\alpha}{2}}\right)^{M},
	\end{equation}
	where $\gamma$ is defined as $\gamma=\sigma^{2}\alpha$.
\end{thm}
\emph{Proof:} We first consider a certain subcarrier $m$.
Since the noise vector $\mathbf{N}(:,m)$ is a Gaussian vector, $\bar{\mathbf{A}}^{H}(:,\Xi_{n}(m))\mathbf{N}(:,m)$ in the case $\mathcal{D}$ should satisfy Gaussian distribution. The mean and the variance of $\bar{\mathbf{A}}^{H}(:,\Xi_{n}(m))\mathbf{N}(:,m)$ are $0$ and $\sigma^{2}\bar{\mathbf{A}}^{H}(:,\Xi_{n}(m))\bar{\mathbf{A}}(:,\Xi_{n}(m))$. We have
\begin{equation}\label{App2}
\begin{aligned}
&\mathrm{Pr}\left\{\left|\bar{\mathbf{A}}^{H}(:,\Xi_{n}(m))\mathbf{N}(:,m)\right|_{2}^{2}\leq\gamma\right\}\\
&\overset{(a)}{=}\mathrm{Pr}\left\{\sigma^{2}\left|\left(\bar{\mathbf{A}}^{H}(:,\Xi_{n}(m))\bar{\mathbf{A}}(:,\Xi_{n}(m))\right)^{\frac{1}{2}}d\right|^{2}\leq\gamma\right\}\\
&=\mathrm{Pr}\left\{\sigma^{2}\left|\bar{\mathbf{A}}^{H}(:,\Xi_{n}(m))\bar{\mathbf{A}}(:,\Xi_{n}(m))\right|\left|d\right|^{2}\leq\gamma\right\}\\
&\overset{(b)}{=}\mathrm{Pr}\left\{\left|d\right|^{2}\leq\frac{\gamma}{\sigma^{2}}\right\},
\end{aligned}
\end{equation}
where $(a)$ comes from defining $d$ as a unit Gaussian variable with mean $0$ and variation $1$, and $(b)$ comes from $\left|\bar{\mathbf{A}}^{H}(:,\Xi_{n}(m))\bar{\mathbf{A}}(:,\Xi_{n}(m))\right|=1$.

To obtain the probability $\mathrm{Pr}\left\{\left|d\right|^{2}\leq\frac{\gamma}{\sigma^{2}}\right\}$, we list a useful lemma \cite{32} as follow.
\begin{thm} \label{lemma5}
	When $d$ is a unit Gaussian variable with mean $0$ and variation $1$, we have
	\begin{equation}\label{App3}
	\mathrm{Pr}\left\{\left|d\right|^{2}\geq r^{2}\right\}\leq 0.8r^{-1}e^{-\frac{r^{2}}{2}}
	\end{equation}
\end{thm}
\emph{Proof:} See \textbf{Lemma 4} in \cite{32}.

Based on (\ref{App3}) in \textbf{Lemma 5}, we have
\begin{equation}\label{App4}
\begin{aligned}
&\mathrm{Pr}\left\{\left|d\right|^{2}\leq\frac{\gamma}{\sigma^{2}}\right\}\\
&=1-\mathrm{Pr}\left\{\left|d\right|^{2}\geq\frac{\gamma}{\sigma^{2}}\right\}\overset{(a)}{\geq}1-0.8\alpha^{-\frac{1}{2}}e^{-\frac{\alpha}{2}},
\end{aligned}
\end{equation}
where $(a)$ comes from (\ref{App3}) and $\gamma=\sigma^{2}\alpha$. Considering that the case $\mathcal{D}$ in (\ref{26}) indicates $\left|\bar{\mathbf{A}}^{H}(:,\Xi_{n}(m))\mathbf{N}(:,m)\right|_{2}^{2}\leq\gamma$ holds for all subcarriers $m=1,2,\cdots,M$, we can obtain
\begin{equation}\label{App5}
\begin{aligned}
&\mathrm{Pr}\left\{\mathcal{D}\right\}\\
&=\mathrm{Pr}\left\{\left|\bar{\mathbf{A}}^{H}(:,\Xi_{n}(m))\mathbf{N}(:,m)\right|_{2}^{2}\leq\gamma,m=1,\cdots,M\right\}\\
&\overset{(a)}{\geq}\left(1-0.8\alpha^{-\frac{1}{2}}e^{-\frac{\alpha}{2}}\right)^{M},
\end{aligned}
\end{equation}
where $(a)$ comes from (\ref{App4}). Therefore, the proof is completed.$\hfill\blacksquare$

\bibliographystyle{gbt7714-numerical}
\bibliography{ref}
\newpage
\biographies
\begin{CCJNLbiography}{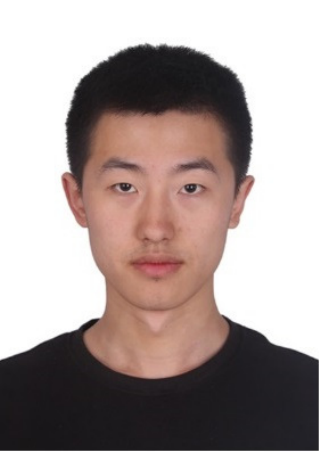}{Jingbo Tan}
(Student Member, IEEE) received his B. S. degree in the Department of Electronic Engineering,  Tsinghua University, Beijing, China, in 2017, where he is currently pursuing his Ph. D. degree. His research interests include precoding and channel estimation in massive MIMO, THz communications, and reconfigurable intelligent surface aided systems. He has received the IEEE Communications Letters Exemplary Reviewer Award in 2018 and the Honorary Mention in the 2019 IEEE ComSoC Student Competition.
\end{CCJNLbiography}

\begin{CCJNLbiography}{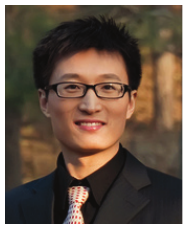}{Linglong Dai}
(Senior Member, IEEE) received the B.S. degree from Zhejiang University, Hangzhou, China, in 2003, the M.S. degree (with the highest honor) from the China Academy of Telecommunications Technology, Beijing, China, in 2006, and the Ph.D. degree (with the highest honor) from Tsinghua University, Beijing, China, in 2011. From 2011 to 2013, he was a Postdoctoral Research Fellow with the Department of Electronic Engineering, Tsinghua University, where he was an Assistant Professor from 2013 to 2016 and has been an Associate Professor since 2016. His current research interests include massive MIMO, reconfigurable intelligent surface (RIS), millimeter-wave/Terahertz communications, and machine learning for wireless communications. He has coauthored the book “MmWave Massive MIMO: A Paradigm for 5G” (Academic Press, 2016). He has authored or coauthored over 60 IEEE journal papers and over 40 IEEE conference papers. He also holds 19 granted patents. He was listed as a Highly Cited Researcher by Clarivate in 2020. He has received five IEEE Best Paper Awards at the IEEE ICC 2013, the IEEE ICC 2014, the IEEE ICC 2017, the IEEE VTC 2017-Fall, and the IEEE ICC 2018. He has also received the Tsinghua University Outstanding Ph.D. Graduate Award in 2011, the Beijing Excellent Doctoral Dissertation Award in 2012, the China National Excellent Doctoral Dissertation Nomination Award in 2013, the URSI Young Scientist Award in 2014, the IEEE Transactions on Broadcasting Best Paper Award in 2015, the Electronics Letters Best Paper Award in 2016, the National Natural Science Foundation of China for Outstanding Young Scholars in 2017, the IEEE ComSoc Asia-Pacific Outstanding Young Researcher Award in 2017, the IEEE ComSoc Asia-Pacific Outstanding Paper Award in 2018, the China Communications Best Paper Award in 2019, and the IEEE Communications Society Leonard G. Abraham Prize in 2020. He is an Area Editor of IEEE Communications Letters, and an Editor of IEEE Transactions on Communications and IEEE Transactions on Vehicular Technology. Particularly, he is dedicated to reproducible research and has made a large amount of simulation code publicly available.
\end{CCJNLbiography}

\end{document}